\documentclass[journal,transmag]{IEEEtran}

\ifCLASSINFOpdf
\else
\fi

\usepackage{times}
\usepackage{epsfig}
\usepackage{graphicx}
\usepackage{float}
\usepackage{subcaption}
\graphicspath{ {./figs/} }
\usepackage{amsmath}
\usepackage{amssymb}
\usepackage{multirow}
\usepackage[]{algorithm2e}
\usepackage{gensymb}
\usepackage[utf8]{inputenc}
\usepackage[english]{babel}
\usepackage{comment}
\usepackage{amsthm}
\usepackage[noadjust]{cite}
\usepackage[usenames, dvipsnames]{color}
\usepackage{mathtools, nccmath}
\usepackage{booktabs}
\usepackage{xr}
\makeatletter
\newcommand*{\addFileDependency}[1]{
  \typeout{(#1)}
  \@addtofilelist{#1}
  \IfFileExists{#1}{}{\typeout{No file #1.}}
}
\makeatother

\newcommand*{\myexternaldocument}[1]{
    \externaldocument{#1}
    \addFileDependency{#1.tex}
    \addFileDependency{#1.aux}
}

\myexternaldocument{supplementary}

\usepackage[switch]{lineno}
 
\usepackage{multirow}
\usepackage[table,xcdraw]{xcolor}
\definecolor{Pink}{rgb}{0.858, 0.188, 0.478}

\DeclarePairedDelimiter{\nint}\lfloor\rceil

\begin{document}
\bstctlcite{IEEEexample:BSTcontrol}

\title{Deep Predictive Motion Tracking in Magnetic Resonance Imaging: Application to Fetal Imaging}

\author{\IEEEauthorblockN{Ayush Singh, 
Seyed Sadegh Mohseni Salehi, 
and
Ali Gholipour
,~\IEEEmembership{Senior Member,~IEEE}}

\thanks{This study was supported in part by the Department of Radiology at Boston Children's Hospital, by the National Institutes of Health (NIH) grants R01 EB018988 and R01 NS106030, and by a Technological Innovations in Neuroscience Award from the McKnight Foundation. The content is solely the responsibility of the authors and does not necessarily represent the official views of the NIH or the McKnight Foundation.

A. Singh and A. Gholipour are with the Department of Radiology, Boston Children's Hospital, and Harvard Medical School, Boston, Massachusetts, USA (email: ayush.singh@childrens.harvard.edu; ali.gholipour@childrens.harvard.edu). S.S.M. Salehi is with Hyperfine Research Inc. (email: sadegh.msalehi@gmail.com). Relevant code can be found at: github.com/bchimagine/DeepPredictiveMotionTracking.

Copyright (c) 2019 IEEE. Personal use of this material is permitted. However, permission to use this material for any other purposes must be obtained from the IEEE by sending a request to pubs-permissions@ieee.org.}}

\maketitle
\begin{abstract}

Fetal magnetic resonance imaging (MRI) is challenged by uncontrollable, large, and irregular fetal movements. It is, therefore, performed through visual monitoring of fetal motion and repeated acquisitions to ensure diagnostic-quality images are acquired. Nevertheless, visual monitoring of fetal motion based on displayed slices, and navigation at the level of stacks-of-slices is inefficient. The current process is highly operator-dependent, increases scanner usage and cost, and significantly increases the length of fetal MRI scans which makes them hard to tolerate for pregnant women. \textcolor{black}{To help build automatic MRI motion tracking and navigation systems to overcome the limitations of the current process and improve fetal imaging, we have developed a new real-time image-based motion tracking method based on deep learning that learns to predict fetal motion directly from acquired images. Our method is based on a recurrent neural network, composed of spatial and temporal encoder-decoders, that infers motion parameters from anatomical features extracted from sequences of acquired slices}. We compared our trained network on held-out test sets (including data with different characteristics, e.g. different fetuses scanned at different ages, and motion trajectories recorded from volunteer subjects) with networks designed for estimation as well as methods adopted to make predictions. \textcolor{black}{The results show that our method outperformed alternative techniques, and achieved real-time performance with average errors of 3.5 and 8 degrees for the estimation and prediction tasks, respectively. Our real-time deep predictive motion tracking technique can be used to assess fetal movements, to guide slice acquisitions, and to build navigation systems for fetal MRI}.   

\end{abstract}


\IEEEdisplaynontitleabstractindextext

\IEEEpeerreviewmaketitle

\section{Introduction}
\subsection{Motivation}
\IEEEPARstart{M}{agnetic} Resonance Imaging (MRI) is a relatively slow imaging technique hence it is extremely susceptible to subject motion. To deal with this limitation, when MRI scans are performed, subjects are instructed to stay completely still. To scan newborns and young children, this requires strategies such as feed-and-wrap, padding, or training, whichever is applicable, to restrain or reduce motion~\cite{malamateniou2013motion,afacan2016evaluation,TryWithout}. There has been extensive research and development in motion-robust sequences and motion correction techniques in MRI (e.g.~\cite{pipe1999motion,thesen2000prospective,maclaren2013prospective,white2010promo,afacan2019retrospective,frost2019markerless,wallace2019head}), \textcolor{black}{however none of these techniques can be universally applied to all MRI sequences and all patient populations. For example none of the above-referenced techniques can be used for motion tracking in fetal MRI, as discussed next}.


Among all rapidly-emerging MRI applications, fetal MRI is, arguably, one of the most challenging, due to uncontrollable, large, and irregular fetal movements~\cite{gholipour2014fetal}. \textcolor{black}{In particular, in mid-gestation fetuses have enough space to stretch and rotate in large angles}. Fetal motion is complex and cannot be monitored or tracked by external sensors or camera systems or accounted for by cardiac and/or respiratory gating. 
Fetal MRI motion correction techniques have thus relied upon \textit{retrospective} image registration solely based on image information~\cite{rousseau2006registration,jiang2007mri,kim2010intersection,gholipour2010robust,kuklisova2012reconstruction,kainz2015fast,alansary2017pvr,marami2017temporal,ebner2018automated}.

Slice-to-volume registration, which has been widely used in retrospective fetal MRI reconstruction, is inherently an ill-posed problem~\cite{ferrante2017slice}. It has a limited capture range as it relies on iterative optimization of intensity-based similarity metrics that are only surrogate measures of alignment between a reference volume and slices. Moreover, a motion-free reference volume may or may not be readily available. To increase capture range, one may use grid search on rotation parameters along with multi-scale registration~\cite{taimouri2015template}; but this approach is computationally expensive as it is based on iterative numerical optimization at test time. For reference volumes, one may use age-matched atlases, e.g.~\cite{gholipour2017normative}, and perform atlas-based registration, e.g.~\cite{taimouri2015template,tourbier2017automated}, however these methods are also computationally expensive for \textit{real-time} application.

To improve capture range and the speed of subject-to-atlas image registration, in a recent work~\cite{salehi2018real2}, deep regression convolutional neural networks (CNNs) were trained to estimate 3D pose of the fetal brain based on image slices and volumes. Partly inspired by~\cite{salehi2018real2}, in this paper we present a novel \textit{deep predictive motion tracking} framework based on long short term memory (LSTM)~\cite{hochreiter1997long} recurrent neural networks (RNNs). While the technique in~\cite{salehi2018real2} addressed \textit{static 3D pose estimation} only (based on regression CNNs), our work here addresses \textit{dynamic, real-time, 3D motion tracking} in MRI, for the first time, using RNNs, exploiting LSTM modules and innovative learning strategies, that are explained in this paper. \textcolor{black}{In static pose estimation we infer 3D pose of the anatomy based on one slice, whereas in dynamic motion tracking, we infer relative pose changes of the subject based on a time series of slices.} Our proposed method, therefore, learns to predict motion trajectory based on MRI slice time series. While motivated by an unmet need in the application domain, our technique was inspired by the most recent advances in computer vision, which are reviewed next, where we also review the related work in fetal MRI and MRI motion tracking.

\subsection{Related Work}

Pose estimation using 2D (digital) images and videos has been extensively researched in computer vision, where algorithms aim to find 3D pose of objects with respect to camera. Work in this area can be studied in two main groups: methods that predict key points leveraging object models to find object orientation, e.g.~\cite{wu2016single}; and methods that predict object pose directly from images to discrete pose space-bins, e.g.~\cite{pavlakos20176},~\cite{tulsiani2015viewpoints} and~\cite{su2015render}. While the majority of pose estimation techniques have been designed as classification methods, the problem has been recently modeled and solved by regression deep neural networks~\cite{mahendran20173d}. Deep CNNs have shown great performance in pose estimation in recent years, e.g.~\cite{mahendran20173d,newell2016stacked,alp2018densepose,andriluka2018posetrack}.

Three-dimensional pose estimation from 3D or stack-of-2D medical images has also been recently addressed using CNNs. For a review of the related pose estimation and registration methods we refer to~\cite{salehi2018real2}. For fetal MRI, in particular, deep regression CNNs were designed for slice-to-volume registration on non-Euclidean manifolds~\cite{hou2017predicting}, and used to estimate transformation parameters for fetal head position to reconstruct fetal brain MRI volumes from slices~\cite{hou20183d}. Real-time fetal head pose estimation was achieved in~\cite{salehi2018real2} by multi-stage loss minimization using mean squared error and geodesic loss, and used for image-to-template and inter-subject rigid registration. 

The above-referenced techniques treat image slices independently. Therefore, while they are powerful in that they learn to predict head position based on single slices (or volumes), they ignore the rich information content of stack of sequentially acquired slices and the dynamics of head motion. Consequently, these methods \textcolor{black}{may be limited in their predictive performance as they ignore (or do not model) the dynamics of motion (e.g., the motion velocity). Moreover, the average 3D pose estimation error of these methods is often high for slices in the boundaries of the anatomy where image features are sparse~\cite{marami2016motion}}. While pose estimation methods can be combined with iterative slice-to-volume registration for head motion tracking, e.g.~\cite{marami2017temporal}; a natural, promising extension of this line of work is dynamic image time series modeling, which has been the subject of our work presented in this paper. \textcolor{black}{In our experiments, we compared our predictive motion tracking technique with zero-velocity and auto-regressive prediction models built upon static 3D pose estimation methods.} 
 

Traditional time series prediction models such as ARIMA (auto-regressive integrated moving average; seasoned, and non-seasoned) expect data to be locally stationary. These are regression models that make strong assumptions about data to predict future values based on past observations. These models shall be paired with other techniques to effectively process and use image time series information; but this integration may not be straightforward. RNNs~\cite{rumelhart1988learning}, on the other hand, can handle non-stationary and nonlinear data. They offer end-to-end framework to take images as input and make predictions, and are flexible in terms of the corresponding objectives.

Variants of RNNs such as networks based on LSTM~\cite{hochreiter1997long} have the capacity to learn the amount of information to remember and forget from past sequences. This makes them less susceptible to unaccounted cases that cannot be easily handled by graph designer of dynamic Bayesian networks (DBNs)~\cite{Gindele2015LearningDB}. Compared to traditional models where 
error propagation leads to error accumulation in long-term prediction, advanced LSTM-based methods, such as sequence-to-sequence (Seq2Seq) learning~\cite{sutskever2014sequence}, can reliably predict variable time steps with long prediction horizons.


Deep predictive motion tracking using RNNs based on video sequences has also been widely studied in robotics and computer vision, e.g.~\cite{sutskever2014sequence,Ondruska2016DeepTS,Krebs2017ASO}. A review of these studies is beyond the scope of this paper, but we briefly review some representative methods and studies. The first group of techniques based on siamese networks detect and use regions close to object locations to track objects, e.g.~\cite{held2016learning,valmadre2017end}. Large datasets can be used to train these networks for feature extraction and region proposals for simultaneous one-shot detection (classification) and online tracking (regression)~\cite{li2018high}. Early performance gains in accuracy were obtained by passing features from an object detector to LSTMs~\cite{ning2017spatially}. In the LSTM category, the Real-time Recurrent Regression (Re$^3$) network~\cite{gordon2018re} combined non-differentiable cropping and warping with feature extraction using a residual network (ResNet), and passed them to LSTM for object tracking. 

\subsection{Contributions}
In this paper we present, for the first time, a dynamic motion tracking framework for MRI based on deep learning. Compared to recent developments in static 3D pose estimation from MRI slices and volumes based on CNNs~\cite{salehi2018real2,hou20183d}, in this work we exploit RNNs for predictive dynamic motion tracking. Compared to motion tracking in computer vision, robotics, digital image and video processing, where 3D pose or projected motion of objects is modeled and estimated based on 2D+time images (videos) with respect to cameras, in this work we deal with 3D rigid motion of anatomy (in the scanner/world coordinate system) from stacks of sequentially acquired slices (3D+time image time series). Consequently, while the majority of human pose tracking or video object tracking methods are formulated and solved as classification problems in a parameter space, we solve a regression problem where 3D rigid motion parameters are estimated based on features directly extracted from MRI time series.

Our contributions are threefold: 1) We developed a learning-based, image-based, real-time dynamic motion tracking in MRI based on deep RNNs:    
Our model encodes motion using LSTM after extracting spatial features from sequences of input images using CNNs, estimates objectives for given images and creates a context vector that is used by LSTM decoders to regress against angle-axis representation and translation offset to predict 3D rigid body motion. The network constitutes multiple representation heads to avoid over-fitting to either rotation or translation parameters. 2) We devised multi-step prediction by feeding output of previous decoder as input to current decoder combined with the context vector. 3) We trained and tested networks on sequences with masked slices that are slices lost due to intermittent fast intra-slice motion.

We developed and tested our method for fetal head motion tracking in fetal MRI, which is a very challenging problem due to the wide range of fetal head positions and motion; but the technique can be used in broader applications. The fetal brain MRI data intrinsically shows a wide feature range due to inter-subject variability and different age of fetuses at the time of MRI scans as well as rapid changes that occur to the fetal brain during gestation. To train and test models we used images of different fetuses scanned at different gestational ages. We simulated motion and also used motion trajectories from sensor recordings of head motion of volunteer subjects to test the generalization capacity of our trained network. We set up a probing task to examine temporal and spatial dependency of our trained model. \textcolor{black}{ Our network infers motion parameters from features extracted from 2D slice time series, therefore it does not require coverage of the entire brain in 3D and hence does not require data that are on a regular grid.}
Our experiments showed that our trained model not only estimated motion trajectories but also was able to make long term predictions based on sequences of fetal brain MRI slices with both simulated and real motion in the test set. The paper is organized as follows: the details of our network and methods are discussed next. Then, the experiments and experimental results are described in Section~\ref{sec:experiments}; which are followed by a discussion in Section~\ref{sec:discussion} \textcolor{black}{and conclusion in Section~\ref{sec:conclusion}}.




\section{Methods}

\subsection{Problem formulation} 
\label{problem_formulation}
Our goal is to take in a sequence of slices $X_1, X_2, ..., X_n$ ($X_n:N\times N$) sampled sequentially (in time) from 3D fetal anatomy (usually acquired in an interleaved manner) in an MRI scan to estimate and predict 3D pose (rotation and slice position) $Y_1, Y_2, ..., Y_{n+m}$ of the fetal brain for current $n$ timesteps as well as future $m$ timesteps \textcolor{black}{(timestep unit defined in Section~\ref{sec:datasets})}. Our technique does not put any restriction on the values of $n$ and $m$. Although $n$ is limited by the number of input slices, $m$ can be variable \textit{i.e.} either less, equal or greater than $n$. \textcolor{black}{The slices are from a stack of sliced anatomy where the anatomy moves in 3D in between slice acquisitions.} For the purpose of this study we assume that the fetal brain is extracted in each slice using a real-time fetal brain MRI segmentation method~\cite{salehi2018real}. For the development and evaluation of predictive motion tracking, we also assume that center-aligned slices are extracted from 3D fetal brain images reconstructed and segmented using the existing techniques~\cite{gholipour2010robust,salehi2017auto}.

\begin{figure}[h]
  \centering
  \includegraphics[keepaspectratio, width=\linewidth]{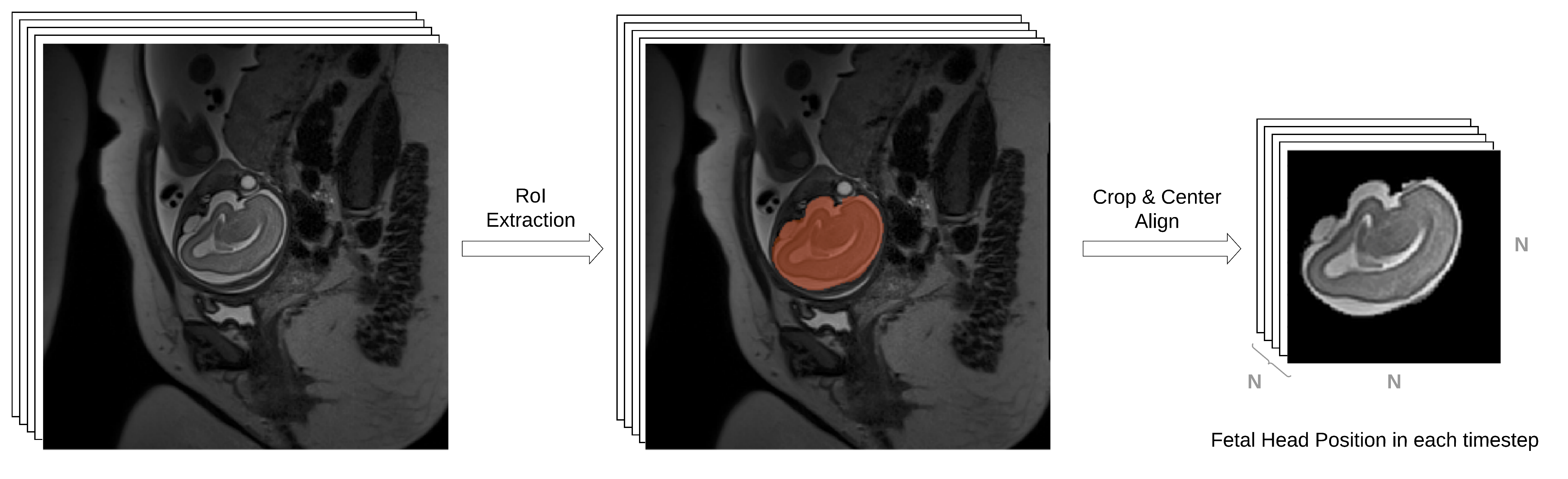}
  \caption{The Region-of-Interest (RoI), here the fetal brain, is extracted using a real-time segmentation technique, e.g.~\cite{salehi2018real}, cropped, center aligned, and intensity normalized to form a volume of stacked slices for deep predictive fetal head motion tracking.}
  \label{fig:data_flow_minimal}
\end{figure}

Figure~\ref{fig:data_flow_minimal} shows how the data is pre-processed and prepared for fetal head motion tracking. The region-of-interest (RoI), which is the fetal brain in this study, is first extracted using a real-time brain extraction method~\cite{salehi2018real} and the slices are cropped, masked, and center-aligned to form a 3D stack. \textcolor{black}{For slices that are corrupted by intra-slice motion (causing full or partial signal loss), the brain extraction method does not generate brain masks that are coherent between those slices and their spatially neighboring slices. The motion-corrupted slices can, therefore, be detected by statistical or learning based methods (e.g., outlier detection~\cite{kuklisova2012reconstruction,marami2016motion} or support vector machines~\cite{marami2017temporal}). Hence,} fetal motion appears as inter-slice motion with occasional black (masked) slices due to intra-slice motion. The problem is \textcolor{black}{formulated as finding 3D rigid transformations, $T$, relative to the starting slice $X_1$, of the fetal head at the times corresponding to slice $X_i$ acquisitions}.

A 3D rigid-body transformation $T$ has 6 degrees-of-freedom represented by a vector $\mathbf{t}$ comprising of three translation ($t_x$, $t_y$, $t_z$) and three rotation \textcolor{black}{$\theta$ }($\theta_x$, $\theta_y$, $\theta_z$) parameters. 
For 3D rotation representation we follow~\cite{salehi2018real2} which uses Euler's theorem and the Rodrigues rotation formula to represent the $3 \times 3$ rotation matrix by the angle-axis representation where the rotation axis is its unit vector and the angle in radians defines its magnitude. Since we center align the images in the pre-processing step, the translation parameters are assumed to be known a priori, which allows us to constrain our parameter space to the slice position $z$ and the rotations $\theta$ represented by the angle-axis formalism. \textcolor{black}{The methods in~\cite{salehi2018real2} can be used to estimate the initial pose and the a priori translation parameters}. 

\subsection{Deep regression RNN for predictive motion tracking}
\begin{figure*}[ht]
  \centering
  \includegraphics[keepaspectratio, width=1.0\textwidth]{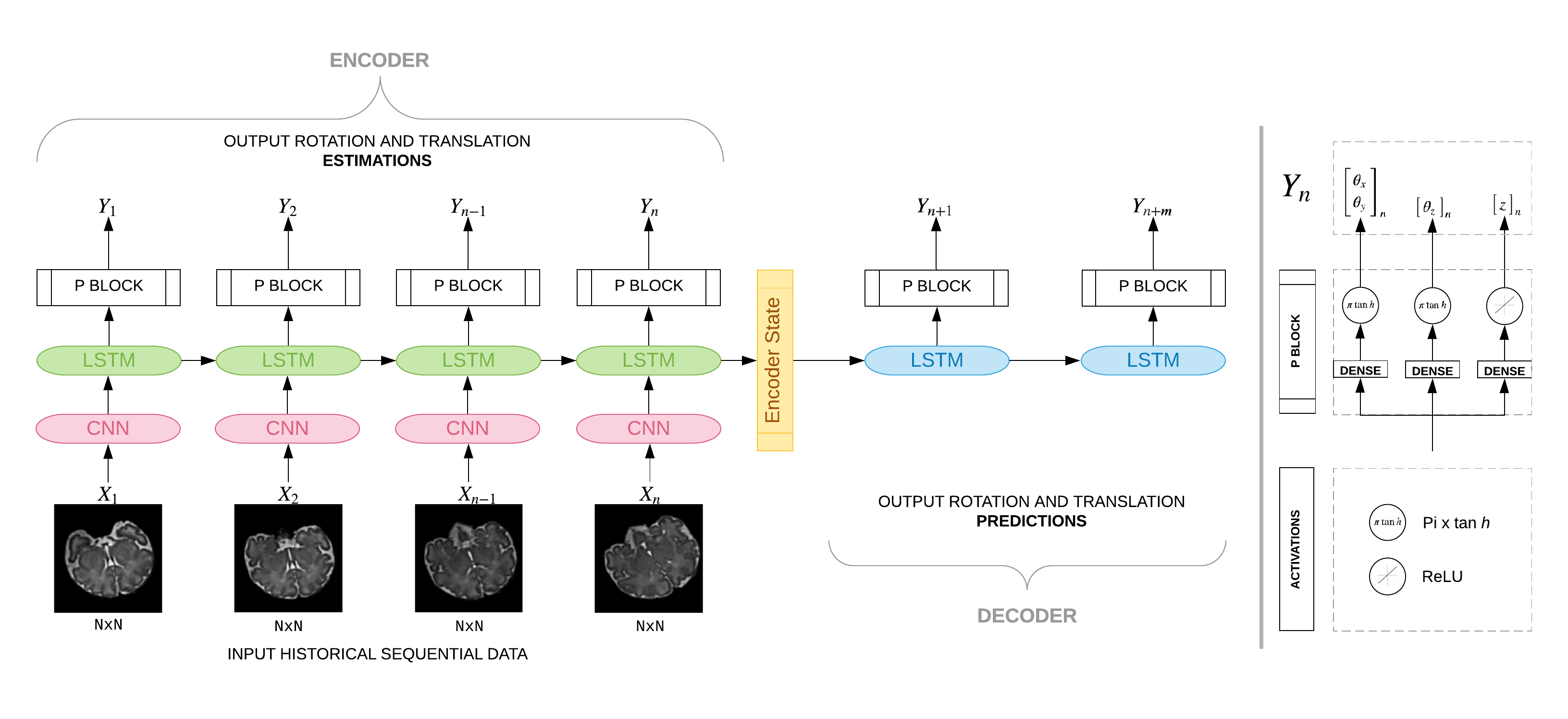}
  \caption{Our many-to-many Seq2Seq model that takes as input sequence of slices and estimates angles as well as predictions. Multiple LSTM units are shown since we unroll our network. All units of the same type and color share weights, hence they get the same gradient update during training. This model comprises of an encoder and a decoder component. The encoder, which contains spatial encoder (CNN) blocks followed by a temporal encoder that contains LSTM units and P blocks, encodes and learns sequence-of-image features to estimate position parameters. The encoder state is fed into the decoder network which comprises of LSTM units followed by P blocks. Each P block has three heads with $\pi \texttt{tanh}$ activation for the rotation parameters and ReLU activation for the slice position.}
  \label{fig:seq2seq}
\end{figure*}

As shown in Figure~\ref{fig:seq2seq}, our deep RNN model for predictive slice-level motion tracking in MRI is built of two main parts: an encoder and a decoder. \textcolor{black}{The encoder network, which is composed of deep CNN blocks followed by unidirectional LSTM and P blocks, takes a sequence of slices $X_1,...X_n$ as input, and estimates a sequence of $n$ transformations as well as an encoder state, which is fed into the decoder network.} Conditioned on the encoder state, the decoder network, which also constitutes LSTM and P blocks, predicts transformations for future time steps $m$. A P block involves three representation heads, each consisting of a dense block and an activation function for regression at the output layer. The activation functions are $\pi \texttt{tanh}$ for the rotation parameters $\theta$ and rectified linear unit (ReLU) for slice position shown here by $z$. In the sections that follow we discuss each of the network components and the details of training.


\subsection{Encoder: Spatial}
\begin{figure}[h]
    \centering
    \includegraphics[width=\linewidth]{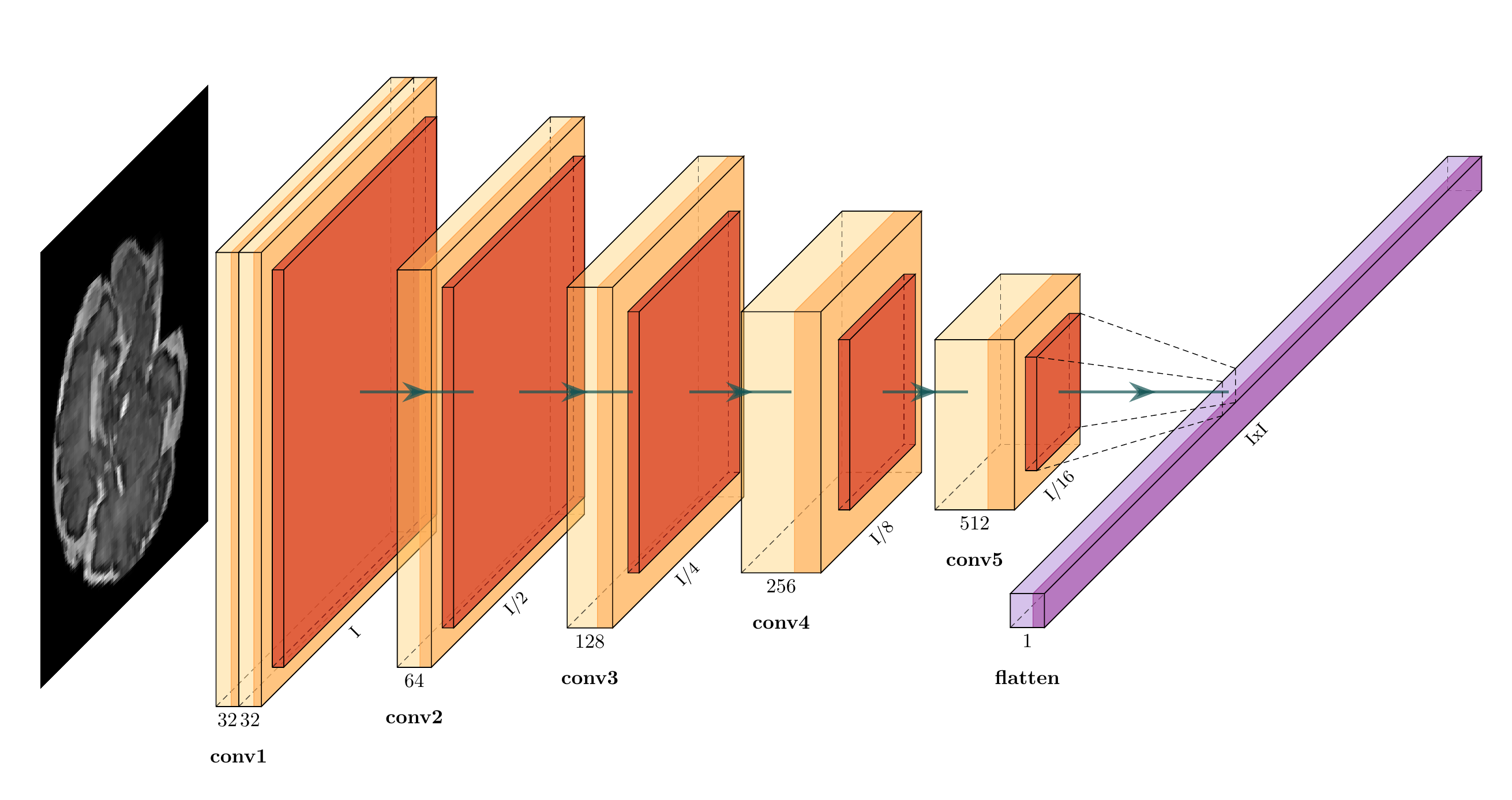}
    \caption{The architecture of the spatial encoder CNN blocks of our deep predictive motion tracking model shown in Figure~\ref{fig:seq2seq}. Each encoder performs $3\times3$ convolutions followed by batch normalization, PReLU~\cite{he2016identity} and \textit{MaxPooling} that down-samples the image in half, extract\textcolor{black}{s} local dependencies and reduc\textcolor{black}{es} computation in downstream layers. This enables fine-grained feature preservation. The number of filters are doubled in each layer until it reaches 512. Finally, \textcolor{black}{features from} the CNN \textcolor{black}{are} flatten\textcolor{black}{ed} and transfer\textcolor{black}{red} as spatial encoding of time step $n$ in the sequence to the LSTM layer of the encoder. \textcolor{black}{Compared to the  deep spatial encoder network used in~\cite{salehi2018real2} to infer 3D pose from a single slice, our CNN is lightweight, which boosts its real-time performance while it effectively encodes features of multiple sequentially-acquired slices and pass them to the LSTM modules to build an encoder state (Figure~\ref{fig:seq2seq})}.}
    \label{fig:cnn}
\end{figure}

For spatial encoding, convolutions ~\cite{lecun1999object} are applied to each slice $X_n$ of a sequence where $n$ is the index of the slice in the sequence. Figure~\ref{fig:cnn} shows the architecture of the spatial encoder network. Through weight sharing the same CNN is trained and applied to all slices. This means there is no dedicated network for each timestep. Instead during training, kernel weights of the same CNN are updated to account for variations in all timesteps. \textcolor{black}{This allows the spatial encoder CNN to learn anatomical variations between different ages, and pass the encoded information into the temporal encoder. We used parametric rectified linear unit (PReLU) as activation function as it has shown better performance than ReLU~\cite{he2016identity}. PReLU avoids the dying ReLU problem, in which a neuron (with ReLU activation) becomes inactive when it gets negative input making the gradient of an inactive neuron zero, hence unable to pass any information via backpropagation.}

\subsection{Encoder: Temporal}
\label{encoder_temporal}
Just as CNN \textcolor{black}{ learns spatial variations, RNN learns variations between elements in a sequence. Since vanilla RNNs face the vanishing gradient problem~\cite{hochreiter2001gradient}, which makes it difficult to propagate gradients back in time}, we used LSTM~\cite{hochreiter1997long}, which also learns what to remember and what to forget. This is important to learn the anatomy and how it is sampled by slices over time using the gating mechanism. Based on encoded image features from the CNN, the LSTM learns to estimate the state of the anatomy, \textit{i.e.} the 3D pose of the anatomy and its sampling. 
LSTM has three primary components: $W, U, b$; where $W$ is the recurrent connection between previous and current hidden layers, $U$ connects inputs to current hidden layer and $b$ is bias:
\begin{equation}
LSTM_{encoder}: X_n, h_{n-1}, c_{n-1} \longrightarrow h_n, c_n
\end{equation}
\begin{equation}
i_n = \sigma(W_i X_n + U_i h_{n-1} + b_i)
\end{equation}
\begin{equation}
f_n = \sigma(W_f X_n + U_f h_{n-1} + b_f)
\end{equation}
\begin{equation}
o_n = \sigma(W_o X_n + U_o h_{n-1} + b_o)
\end{equation}
\begin{equation}
\hat{c}_n = tanh(W_{\hat{c}} X_n + U_{\hat{c}} h_{n-1} + b_{\hat{c}})
\end{equation}
\begin{equation}
c_n = f_n \odot c_{n-1} + i_n \odot \hat{c}_n
\end{equation}
\begin{equation}  \label{eq:7}
h_n = o_n \odot tanh(c_n)
\end{equation}

For each time step $n$, the memory cell $c_n \in \mathbb{R}^n$ is called as it controls exposure of the previous memory $c_{n-1}$ with current input $X_n$. This is done by combining $c_{n-1}$ multiplied by the forget gate $f_n$, with the computed hidden state $h_n$ multiplied by the input gate $i_n$. These are called gates because they squash values between 0 and 1 using the sigmoid activation function $\sigma$. The element-wise multiplication $\odot$ controls how much of information is let through: The input gate controls how much of the current input goes through; the forget gate controls the throughput of the previous state; and the output gate controls the amount of exposure of the internal states to the next timesteps (or the downstream layers). All gates have dimensions equal to that of the hidden layer $h_n$, which is computed by multiplying the hyperbolic tangent $tanh$ of memory $c_n$ with the output $o_n$. $\hat{c}_n$ is the candidate hidden state that connects the current input $X_n$ to the previous hidden state. One can ignore old memory completely (all zeros $f_n$) or ignore states (all zeros in $i_n$), but we chose to store nuances of changes in data over time thus the values were chosen to be between 0 and 1.

Flattened feature maps pass from the spatial encoder to the unidirectional LSTM network. Output of each time step of the encoder and decoder LSTM go through dense fully-connected layers to get estimated and predicted parameters. The last non-linear function with weights $W_{\theta_{xyz}}$ on top of the dense layer is $\pi\times\tanh$ which limits the output of each element from $-\pi$ to $+\pi$ and simulates the constraints of each element of the rotation vector ($\theta_x, \theta_y$) and $\theta_z$ independently:
\begin{equation}  \label{eq:8}
\theta^{xyz}_n = \pi tanh(W_{\theta_{xyz}} o_n + b_{\theta_{xyz}})
\end{equation}
The slice index ($z$) estimator head with weights $W_{z}$ contains a scalar, as the network tries to estimate the continuous slice index along with its orientation. For inference, the continuous index is rounded (i.e. $\nint{z}$) to infer a discrete slice number.
\begin{equation} \label{eq:9}
z_n = max(0, W_{z} o_n + b_{z}) \quad (ReLU)
\end{equation}

\subsection{Decoder: Modeling variable and long term predictions}
\label{decoder}
The conventional approach to predict sequential data is to use $n$ steps of the sequence from the past to predict the immediate future time step $n+1$ and repeat recursively to make future predictions up until the desired prediction horizon. This model, however, shows limited multi-step prediction performance in applications such as image-based motion tracking as it faces issues raised by compounding errors especially when initial predictions may exhibit relatively large amounts of error. To mitigate this issue and make variable-length, long-term predictions we follow the idea of sequence to sequence learning~\cite{sutskever2014sequence}. 
In this approach, an LSTM encodes the input sequence of images into a fixed dimension vector, and another LSTM decodes the target sequence from this vector. The advantage of this technique is that we no longer need to rely on encoder estimates to predict variable-length time steps of the future as encoder and decoder are two separate LSTM networks. Figure~\ref{fig:seq2seq} shows our LSTM network unrolled. 

Each decoder is trained to predict parameters of the following step.
Therefore, input to the first decoder is the estimation vector $\hat{Y}_n$ 
of the last slice $X_n$ from the encoder and the rest of the decoder takes output of the previous decoding step $\hat{Y}_{m-1}$ so that over time the model learns to correct its own mistakes. 
\begin{equation} \label{eq:10}
LSTM_{decoder}: \hat{Y}_{n+m-1}, h_{n+m-1}, c_{n+m-1} \longrightarrow h_{n+m}, c_{n+m}
\end{equation}

The goal of decoding is to model the conditional probability of $P(Y_1,..,Y_{n+m}|X_1,..,X_n)$. The decoder uses $h_n, c_n$ from encoder as its initial state to compute $P(Y_{n+m})$. 
However the decoder does not directly model $P(Y|X)$, its power comes from modeling probability of current output 
with respect to all previous timesteps $P(Y_{n+m}|Y_{<n+m}, X_n)$ where $Y_{<n+m}$ represents output from $1$ to ${n+m-1}$. The posterior probability of the output state given inputs, with model parameters $\gamma$, is as follows
\begin{equation}  \label{eq:11}
P_\gamma(Y|X) = \displaystyle\prod_{n=1}^{n+m} P_\gamma(Y_n|Y_{<n}, X)
\end{equation}


\subsection{\textcolor{black}{Loss functions}}
\label{split_multiple_heads}
The coupling between in-plane and out-of-plane rotation with the slice select direction and slice location $z$ hinders optimization and learning~\cite{hou2017predicting}. To alleviate this issue, we divided the rotation $\theta$ regression heads from Equation (\ref{eq:8}) and added a hidden layer one each for $\theta_{xy}$ and $\theta_z$ as follows:
\begin{equation} \label{eq: 12}
    \theta^{xy}_n = \pi tanh(W_{\theta_{xy}} tanh(W_{\theta_{xyz}} o_n + b_{\theta_{xyz}}) + b_{\theta_{xy}})
\end{equation}
\begin{equation}  \label{eq:13}
    \theta^{z}_n = \pi tanh(W_{\theta_{z}} tanh(W_{\theta_{xyz}} o_n + b_{\theta_{xyz}}) + b_{\theta_{z}})
\end{equation}
which changes our loss calculation from 
\begin{equation}  \label{eq:14}
L_{Total} = L_{\theta_{xyz}} + L_{z}
\end{equation}
to
\begin{equation}  \label{eq:15}
L_{Total} = L_{\theta_{xy}} + L_{\theta_z} + L_{z}
\end{equation}
For training, we minimized \textcolor{black}{mean squared error (MSE)} for both estimation and prediction $L_{Total} = L_{estimation} + L_{prediction}$ where $L = \Vert Y - \hat{Y} \Vert_2$.
We used $tanh$ as activation of this hidden layer as its derivative provides a stronger gradient for regression tasks compared to $ReLU$ or $sigmoid$ functions. \textcolor{black}{In summary, we split our rotation loss in two separate layers; and regressed our rotation and slice location parameters using the backpropagation algorithm.}

\section{Experiments}
\label{sec:experiments}
To train, test, and evaluate our method we conducted experiments with real fetal MRI data with simulated motion and motion tracking data of volunteers who moved inside the scanner while motion parameters were recorded using an external motion tracking sensor. \textcolor{black}{All fetal MRI and volunteer experiments were performed under protocols approved by the institutional review board committee, and written informed consent was obtained from all pregnant women volunteers and other volunteers}. We divided our main experiments into estimation for 10 timesteps and prediction for 10 timesteps; and evaluated our trained model for generalization, robustness, and latency; and compared our results against pose estimation networks in particular those based on SVRNet~\cite{hou2017predicting}, PoseNet~\cite{salehi2018real2}, and our baseline models for estimation and prediction. \textcolor{black}{Further, we tested our estimated motion parameters with a retrospective slice-to-volume reconstruction method~\cite{ebner2020automated}}. In this section, we describe the fetal MRI data and its pre-processing first; and then the details of our experiments that involved generating the training data and the results of estimation and prediction for both simulated and real motion trajectories.

\subsection{Fetal MRI dataset}
\label{sec:datasets}

The fetal MRI dataset consisted of \textcolor{black}{repeated multi-planar T2-weighted single shot fast spin echo scans as well as reconstructed T2-weighted fetal brain MRI scans of 82} fetuses scanned at a gestational age (GA) between 21 and 37 weeks (mean=30.1, stdev=4.6) on 3-Tesla Siemens Skyra scanners with 18-channel body matrix and spine coils. \textcolor{black}{The in-plane spatial resolution of the original scans was 1 \textcolor{black}{mm}, the slice thickness was 2-3 \textcolor{black}{mm}, and the temporal resolution for slice acquisition was equal to the repetition time (TR), which was 1.5s. This defined the time unit for slice-level motion tracking, so the timestep in motion tracking was 1.5s. Brain masks were automatically extracted on slices of the original scans using} the real-time algorithm in~\cite{salehi2018real}. The scans were automatically cropped around the fetal head RoI (based on the masks) and were then processed using slice-by-slice motion correction to reconstruct a super-resolved 3D volume~\cite{gholipour2010robust,kainz2015fast} \textcolor{black}{at an isotropic resolution of 0.8 \textcolor{black}{mm}}. Final 3D brain masks were then generated on the reconstructed images using Auto-Net~\cite{salehi2017auto} and manually corrected in ITK-SNAP~\cite{yushkevich2006user} as needed.

Brain-extracted reconstructed volumes were then registered to a spatiotemporal fetal brain MRI atlas described in~\cite{gholipour2017normative}. We normalized the intensity of the reconstructed images to zero-mean, unit-variance. The set of \textcolor{black}{82 scans were split into 30, 6, 40 and 6 for training, validation, test, and reconstruction, respectively; where the GA range spanned over 29 to 35 weeks for the training set, and from 26 to 37 weeks for the test set. We intentionally chose a narrower age range for the training set than the test set to examine the generalization capacity of the trained models \textcolor{black}{on extended age ranges}. To generalize well, the trained models had to account for both intrinsic inter-subject anatomical variations (due to different fetuses in the training and test sets) and anatomical variations due to different maturation levels of fetuses scanned at different GA ranges. The training, validation, test, and reconstruction set splits never had scans of the same subject. The GA of the reconstruction set subjects were 28, 30, 32, 32, 35, and 37; and between 6-10 (mean=7) multi-plane stacks of slices were used to reconstruct a volume for each of those subjects.} 


\subsection{Generating the Training Data}
\label{sec:trainset}
To achieve our goal of predicting motion and slice position from sequences of slices, we aimed to train networks to learn the patterns of slice sampling and fetal head motion in reference to the fetal brain anatomy while it develops during gestation. To generate the training, validation, and test data for this purpose, from the pre-processed fetal MRI data, we generated sequences of fetal MRI slices with motion. This involved two sampling components: spatial sampling of slices and temporal sampling of spatial slices to model fetal motion.
For slice excitation and spatial sampling, we sampled sequentially along permuted $Z$ axes with 5 \textcolor{black}{mm} slice gap to account for fetal MRI acquisitions that are interleaved.

\begin{figure}[h]
    \centering
    \includegraphics[keepaspectratio, width=\linewidth]{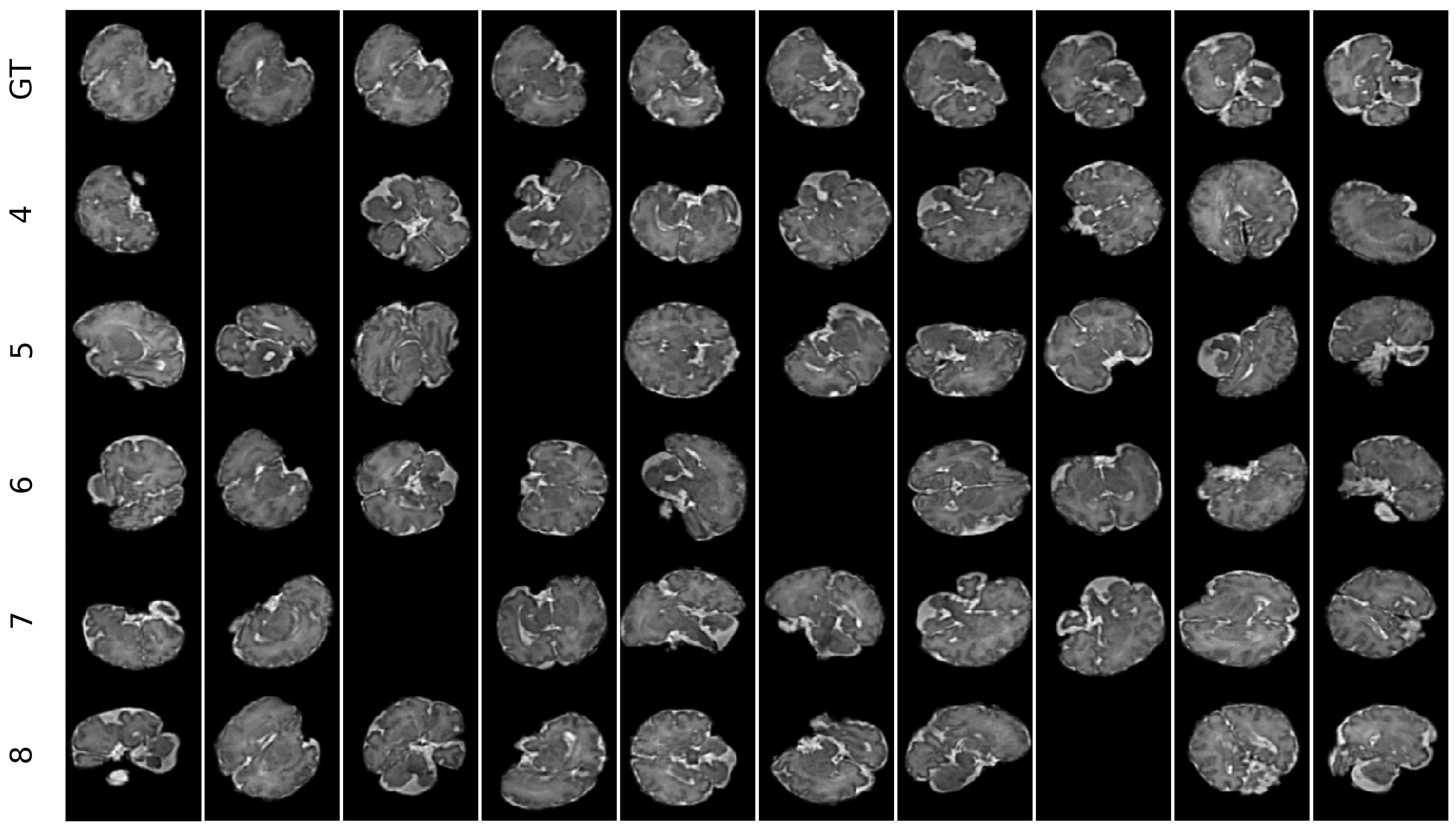}
    \caption{A demo of five sequences of 10 timesteps each generated with different speeds of motion (corresponding to the number of \textcolor{black}{spline} control points from 4 to 8) from the 3D reconstructed fetal brain MRI scan of GA 35 weeks (shown at the top row). Randomly masked slices indicate slices corrupted by intra-slice motion.}
    \label{fig:gen_seq_example}
\end{figure}

For temporal sampling to generate dynamic transformations corresponding to fetal motion, we exploited curve fitting with smoothing cubic Splines for each of the rotation angles. In this scheme, smoothing cubic splines generated different motion trajectories by interpolating curves between randomly-generated control points. The number of control points varied to control speed of motion. This was analogous to how fast or slow the fetus moved between scans. 
Further, to account for fast motion that disrupts slice encoding, we randomly masked a timestep in all slices. This resembled intra-slice motion as the brain masking technique in~\cite{salehi2018real} generated all-zero masks for motion-corrupted slices. Figure~\ref{fig:gen_seq_example} shows five 10-timestep sequences generated from the reference (GT) image sequence with random patterns and different speeds of motion.

We sampled 32 sequences for each subject in the training set 300 times (epochs). This led to 30 subjects $\times$ 32 sequences (1 batch size of 5 speed categories) $\times$ 300 times = 288,000 sequences for training, where speed of motion was controlled by the number of smoothing spline control points sampled from a normal distribution ($\mu=6.4$, $\sigma=1.36$, bounds=$[4,8]$). The initial rotation matrices were bound to [$-\pi/3,\pi/3$] range; and the rotation parameters $\theta_x, \theta_y, \theta_z$ were sampled from a zero-mean normal distribution in the [$-\pi/6,\pi/6$] range. \textcolor{black}{This led to maximum rotation bounds of [$-\pi/2,\pi/2$]. For the validation set we followed the same sampling strategy, which led to $6 \times 32 \times 300 = 57,600$ sequences for validation}.

\subsection{Test Datasets}
\label{sec:testset}
To test and compare algorithms, we sampled 32 sequences per speed of $[4,8]$ where we followed the spatial and temporal sampling strategies described in the previous section. This resulted in a total of 40 test subjects $\times$ 32 samples = 1280 sequences of \textcolor{black}{20 timesteps (10 estimation + 10 prediction)} each for test. Even though our main goal was to evaluate one-step ahead prediction, having 10 prediction timesteps allowed us to test efficacy of the model on long-term predictions. 
While our training data was limited to sequences generated from fetal MRI scans using the described procedure, to evaluate the generalization capacity of the trained models for new (unseen) patterns of motion, in addition to the test set described above, we used motion data recorded using head motion tracking sensors~\cite{gholipour2011motion} from 10 volunteers. Rigid 3D transformation parameters were recorded in the scanner as volunteers moved their head with different patterns and speeds during scans. We applied these motion trajectory parameters to each of the 40 fetal test subjects, which led to a total of 400 new sequences with realistic motion patterns that differed from the motion patterns of the training data. \textcolor{black}{The scans of the 6 test subjects in the reconstruction set were directly used in the reconstruction experiments. The details of the implementation and experiments are discussed next.}



\subsection{Implementation and Experimental Details}
We used the mean square error (MSE) loss and the RMS-prop optimizer with initial learning rate of 0.001 ending in 0.00001 over the course of 300 epochs, decreasing the learning rate when the loss plateaued for 50 consecutive epochs. Due to the temporal nature of MRI slice acquisitions and the fact that the boundary slices did not include sufficient anatomical features, we limited estimation and prediction of motion trajectories to slices $s_i ; i \in [0.4S, 0.9S]$, where $S$ was the total number of slices in each reconstructed brain volume.

We conducted experiments and evaluated our model in both estimation and prediction tasks. For estimation, we compared our model (with 4.7M parameters) with two state-of-the-art fetal MRI pose estimation methods, \textit{i.e.} an 18-layer residual network (ResNet) with two regression heads, one for angles $\theta$ and the other for slice location $z$, based on PoseNet~\cite{salehi2018real2} (with 11M parameters), and a VGG16-style network based on SVRNet~\cite{hou20183d} (with 14.7M parameters). 
Since SVRNet chose VGG16 among several other models, namely GoogLeNet, CaffeNet, Inception v4, and ResNet, we only compared against VGG16, as according to~\cite{hou20183d} it generated the lowest MSE.

For prediction, we conducted experiments for one-step and multi-step ahead predictions. To implicitly model motion states (i.e. to estimate motion velocity and acceleration) we needed a window size of at least three timesteps. In our experiments we used a window size of 10 for estimation and prediction \textcolor{black}{each}. For multi-step prediction, we limited our evaluation to 10 timesteps in the future although this was a choice and not a theoretical limit on the prediction horizon. We compared our predictor against three baselines: 1) a naive predictor that used estimation at current time as one-step ahead prediction (referred to as zero velocity predictor); 2) an auto-regressive model that recursively used its own predictions in a sliding window of size 10 to predict multi-step motion trajectories; and 3) a predictive model that we adopted based on the network proposed in~\cite{martinez2017human}. In this model (with 44M parameters), the data was passed directly into an LSTM without spatial feature encoding, thus we refer to it as directLSTM. 

\textcolor{black}{For the volume reconstruction experiments from multiple scans, we rearranged slices of the original fetal MRI scans (with inter-slice motion) based on slice timing, estimated 3D pose, and passed estimated parameters from our motion tracking algorithm along with volume-to-volume transformation to the canonical atlas space~\cite{salehi2018real2} to NiftyMIC~\cite{ebner2018automated,ebner2020automated}. We then compared the reconstructions to reconstructions directly performed by NiftyMIC in the atlas space. We compared reconstructed images using Structural Similarity Index (SSIM) which has a range of -1 to +1 where 1 means a perfect match, and Normalized Root Mean Square Error (NRMSE), which ranges between 0 and 1 where 0 means perfect match (0 error).}

\subsection{Results}

\begin{figure*}
    \centering
    \includegraphics[keepaspectratio, width=1.0\textwidth]{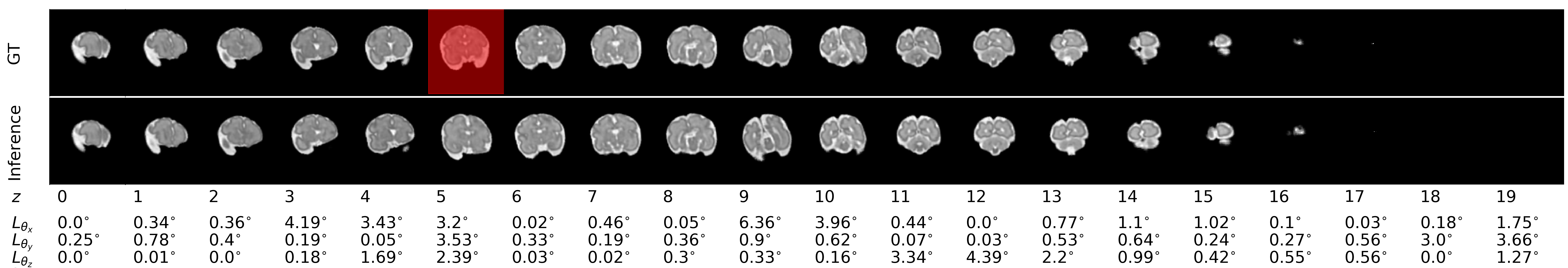}
    \includegraphics[keepaspectratio, width=1.0\textwidth]{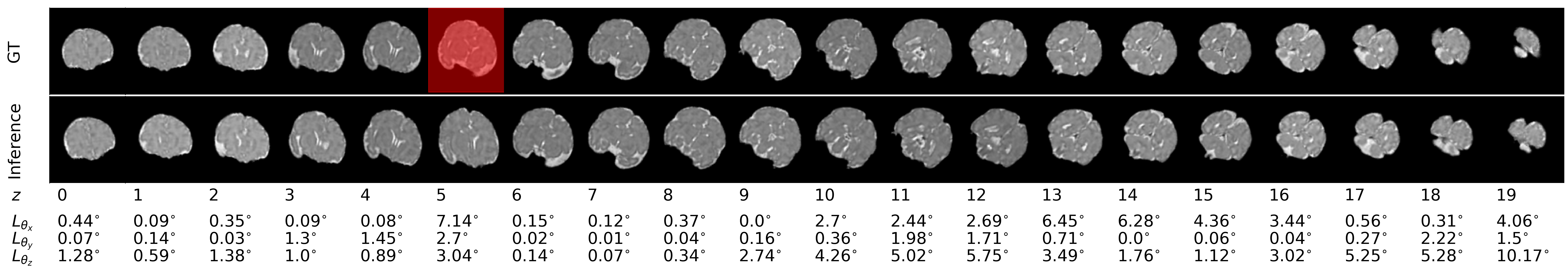}
    \caption{Inference (i.e. estimation for the first 10 timesteps and prediction for the rest of the 10 timesteps) in the bottom rows has been compared to the ground truth sequence in the top rows for scans of two fetuses: the first figure is a scan of a 28-week, and the second figure is a scan of a 36-week GA fetus from the test set. \textcolor{black}{Errors based on the MSE loss (Section~\ref{split_multiple_heads}) have been shown underneath each timestep.} In these figures the slices shown with red masks were masked in the input sequence. It can be seen that the estimated slices (in the bottom rows) corresponding to the masked slices, showed relatively larger error, but the masked slices did not have a major effect on predictions. Slight increase in prediction error with prediction time horizon was seen in the test sequence, but the predictions were overall accurate.} 
    \label{fig:seq_estimation}
\end{figure*}

\begin{figure}[h]
    \centering
    \includegraphics[keepaspectratio, width=\linewidth]{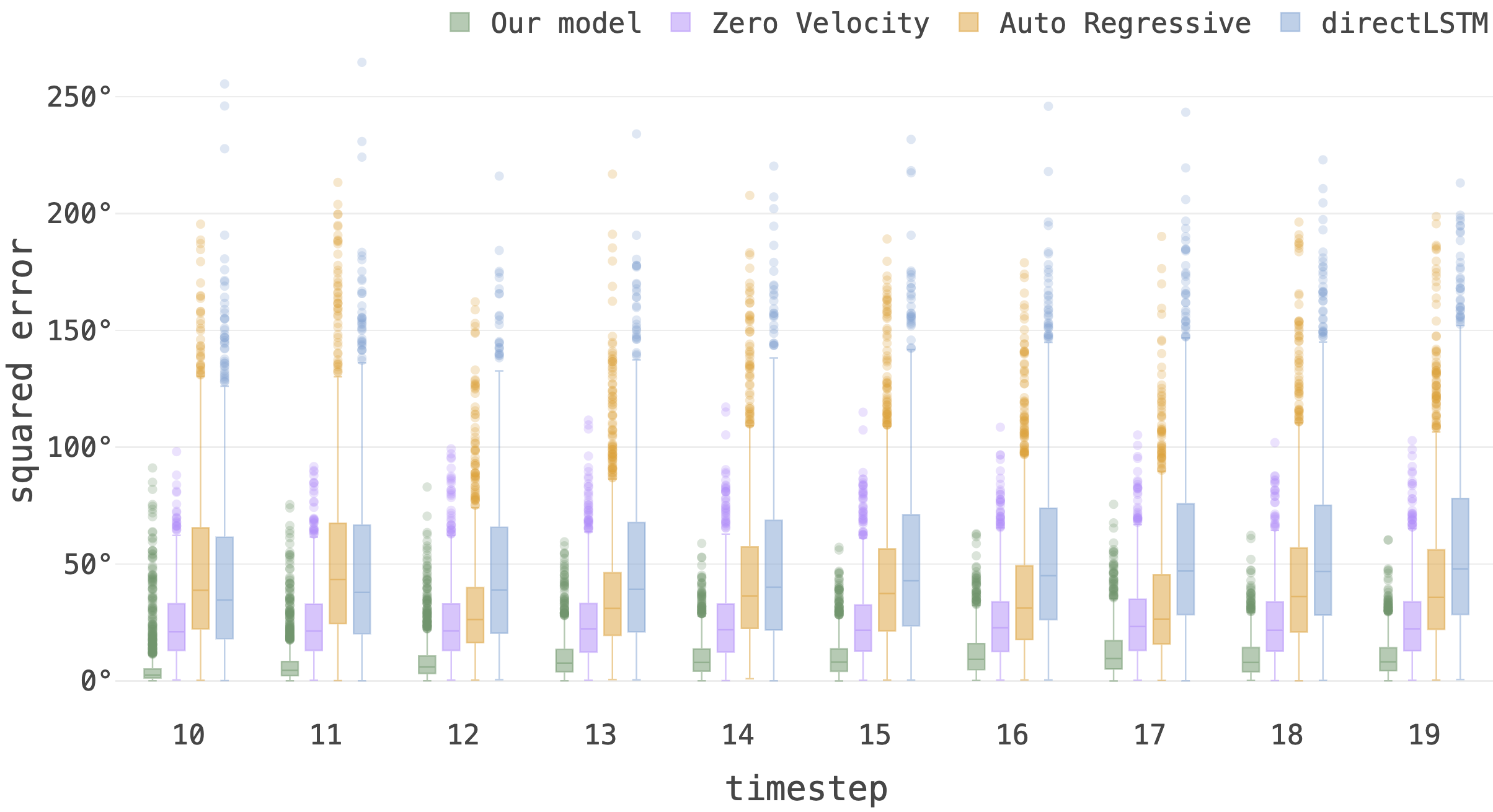}
    \caption{Boxplots showing the statistics of the \textcolor{black}{average rotational MSE loss on test data} computed for prediction per timestep. Our model outperformed all other prediction models implemented and tested here (i.e., zero velocity, auto-regressive, and directLSTM).} 
    \label{fig:pred_loss}
\end{figure}

Figure~\ref{fig:seq_estimation} shows 10 estimated and 10 predicted timesteps for a train case and a test case compared to the ground truth slices in the top rows. The predicted rotation was accurate after multiple timesteps. \textcolor{black}{Table~\ref{total_loss_table} shows average loss of estimation and prediction tasks (defined in Section~\ref{split_multiple_heads}) on the test data with synthetic motion, along with the standard errors computed between groups of fetuses in the test set based on the prediction timestep (time), age at scan, and speed of motion, for the ablation studies as well as the comparisons to baseline and alternative methods. We compared our "full model" trained with sequences with masked slices (resembling slices corrupted by intra-slice motion) and split loss explained in Section~\ref{split_multiple_heads} against our "baseline" which was trained without masked slices in the training set sequences and without split heads, and "masked bl" which was trained with masked slices but without the split loss functions.} The best results in each comparison, shown in bold, show that our full model outperformed the baselines and all other models in both estimation and prediction tasks. The low standard errors of our model show its consistent and robust performance \textcolor{black}{with respect to the different characteristics of the test data. Figure~\ref{fig:pred_loss} shows the rotational MSE of multi-step prediction per timestep (estimation for time 10 and predictions for times 11 to 19) on test data, where the images corresponding to time points 1 to 10 were the inputs to the model}. 

\begin{table}
\centering
\begin{tabular}{llllll} 
Model            & $\mu$ error & $\sigma_\mu$ & $\sigma_\mu$ time & $\sigma_\mu$ age & $\sigma_\mu$ speed  \\ 
\midrule\midrule
VGG16                & 129.33              & 11.74            & 3.72                    & 3.48              & 9.51           \\
Resnet18             & 82.60               & 5.76             & 3.55                    & 1.31              & 3.34           \\
Our baseline         & 20.19               & 2.57             & 1.21                    & 2.23              & 2.06           \\
Our masked bl.       & 9.10                & 2.31             & 1.11                    & 1.92              & 2.45           \\
Our full model       & \textbf{3.55}       & \textbf{0.22}    & \textbf{0.17}           & \textbf{0.05}     & \textbf{0.23}  \\
\midrule
directLSTM           & 103.20              & 3.09             & 0.97                     & 13.52                & 5.80           \\
Zero velocity        & 74.14               & 1.09             & 0.86                     & 1.77                 & 1.32           \\
Auto regressive      & 96.77               & 1.66             & 0.69                     & 1.83                 & 2.17           \\
Our baseline         & 33.51               & 2.35             & 1.17                     & 1.23                 & 1.11  \\
Our masked bl.       & 11.28               & 1.28             & 1.17                     & 0.23                 & \textbf{0.51}  \\
Our full model       & \textbf{8.07}       & \textbf{0.72}    & \textbf{0.42}            & \textbf{0.39}        & 0.59  \\
\end{tabular}
\caption{Mean squared error ($\mu$ error) for estimation and prediction \textcolor{black}{of 3D pose in degrees} along with the overall standard error of mean ($\sigma_\mu$) and the standard error of different timesteps, ages, and speed of motion for the test data. The top part of the table compares estimation models and the bottom part compares prediction models. In these comparisons we also tested our model trained without any masked slices 
in the sequences, referred to as the ``baseline", our second baseline trained with masked slice sequences but without the split heads and the loss function explained in Section~\ref{split_multiple_heads} (referred to as "masked bl.") and our "full model" trained with both masked slices and the split loss function. Significant reduction in both estimation and prediction errors was achieved by our full trained model compared to baselines and all other compared models. Low standard errors show that our model performed consistently, and was robust to variations in data, timesteps, GA, and the speed of motion.} 
\label{total_loss_table}
\end{table}

\begin{figure}[h]
    \centering
    \includegraphics[keepaspectratio, width=\linewidth]{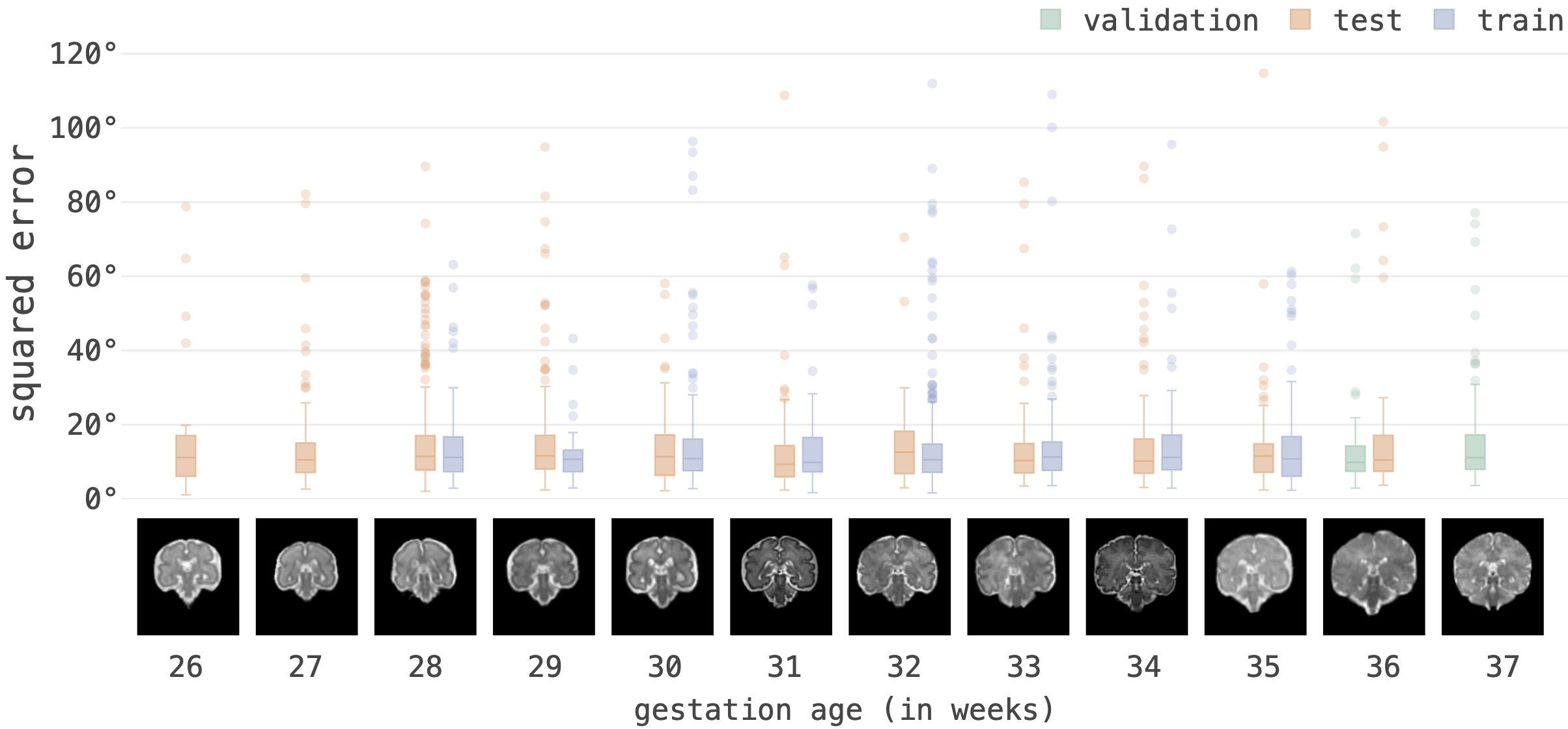}
    \caption{\textcolor{black}{Average MSE of 3D pose in degrees} of one-step ahead prediction tasks for 10 timesteps grouped by GA. Consistent errors show that our model generalized well to variations in anatomy and GA outside of the domain and range that it was trained on.}
    \label{fig:pred_compare}
\end{figure}

In the next sets of experiments, we evaluated our model for 1) its generalization performance for our test data that included subjects scanned at gestational ages not included in the training set; 2) its performance for different speeds of motion; 3) its robustness in the presence of intra-slice motion (i.e. lost slices in the input sequence due to fast motion that disrupted signal during slice encoding); and 4) its generalization and robustness to motion patterns that were different from the motion patterns in the training data (i.e. motion patterns recorded from volunteer subject experiments). \textcolor{black}{Figure~\ref{fig:pred_compare} shows boxplots of the MSE of the estimation and prediction tasks for 10 timesteps grouped by gestational age and datasets.} The consistency in error statistics across test and train datasets and GA, indicate that the trained model was robust and generalized well to the test data.




\textcolor{black}{Table~\ref{mask-table} shows the MSE of pose estimation, one-step and multi-step prediction for test data grouped by the location of a lost slice (due to intra-slice motion) in the input slice sequence}. This table compares the performance of two models: our model trained without any missed (masked) slices in the training sequences (referred here as the baseline); and our full model trained with randomly missed (masked) slices in the training set. These results show that 1) in the baseline model, both estimation and prediction errors were higher when the lost slice was closer to the end of the input sequence; \textit{i.e.} missing slice 10 in the sequence led to much higher errors (shown in red) compared to missing slices in earlier locations; 2) Our full model performed better than the baseline with much more consistent and robust performance; and 3) Our full model's performance degrades if the first timestep is masked because being first timestep it does not have information from past and by masking it. These show that when our model was trained with randomly masked slices in the training sequences, it learned to rely less on the last slices in the sequence to gain robustness to intra-slice motion. 

\begin{table}[h]
\centering
\begin{tabular}{l|lll|lll}
\multirow{2}{*}{\textbf{\begin{tabular}[c]{@{}l@{}}Timestep\\ Masked\end{tabular}}} & \multicolumn{3}{c}{\textbf{Baseline model error}} & \multicolumn{3}{c}{\textbf{Masked model error}} \\
  & \textbf{Est}  & \textbf{OSP} & \textbf{MSP} & \textbf{Est}  & \textbf{OSP} & \textbf{MSP} \\\hline
\textbf{No Mask} & \textbf{1.37} & \textbf{2.97} & \textbf{7.41} & \textbf{1.03} & \textbf{2.93} & \textbf{7.69} \\
1 & 5.83 & 4.42 & 10.48 & 4.86 & 3.70 & 10.05 \\
2 & 4.83 & 3.03 & 7.58 & 2.97 & 2.87 & 7.62 \\
3 & 4.36 & 2.98 & 7.50 & 2.17 & 2.86 & 7.61 \\
4 & 3.06 & 3.03 & 7.98 & 1.87 & 2.71 & 7.98 \\
5 & 3.87 & 3.05 & 8.13 & 2.01 & 2.83 & 7.41 \\
6 & 3.29 & 4.06 & 8.39 & 2.43 & 2.91 & 7.63 \\
7 & 3.25 & 4.17 & 8.65 & 2.59 & 2.93 & 7.69 \\
8 & 3.91 & 6.37 & 9.21 & 2.61 & 3.02 & 7.74 \\
9 & 4.06 & 6.78 & 10.74 & 2.68 & 3.59 & 8.15 \\
10 & 4.19 & \leavevmode\color{red}17.37 & \leavevmode\color{red}15.89 & 3.88 & 6.96 & 9.54
\end{tabular}
\caption{Results of a probing task on our full model trained with masked data against our model trained on unmasked data (baseline): \textcolor{black}{3D pose MSE in degrees of estimation (Est), one step prediction (OSP) and multi-step prediction (MSP) on test data, which} have been shown based on the timestep in which a slice was masked in the test sequence (first column). Results of both models on unmasked test data (first row) were similar, however the prediction performance of the baseline model indicates that to make predictions this model put a heavy weight on slices that appeared towards the end of the sequence. On the other hand, our full model trained with randomly-masked sequences, performed more consistently and robustly with respect to the position of the masked slice in the input test sequence.} 
\label{mask-table}
\end{table}



\begin{figure}[h]
    \centering
    \includegraphics[keepaspectratio, width=\linewidth]{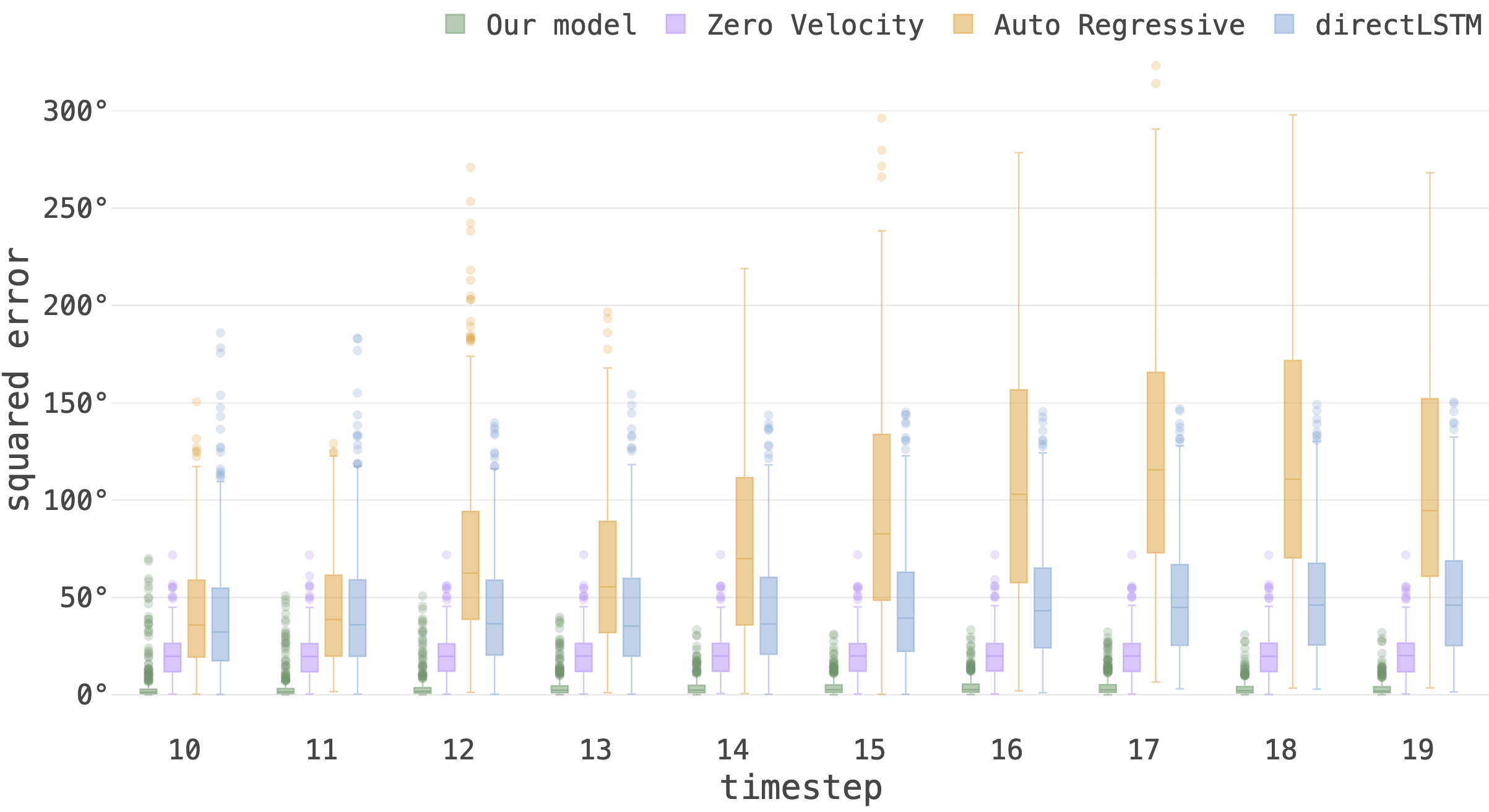}
    \caption{\textcolor{black}{3D pose MSE in degrees of multi-step prediction for the test data with motion trajectories recorded} from volunteers, shows the generalization capacity of our model on real motion patterns. In all baseline models, the prediction error increased with the prediction steps due to compounding errors. 
    In contrast, by passing context from encoder and prediction from previous decoding, our model maintained low error in multi-step prediction.}
    \label{fig:volunter_prediction}
\end{figure}

We evaluated the generalization capacity \textcolor{black}{of our model trained on data with synthetic motion, on motion trajectories recorded from volunteer subjects (that were never used in training).}  Figure~\ref{fig:volunter_prediction} shows the mean squared pose prediction error for different timesteps for the test data with the recorded motion trajectories, obtained from our full model and other predictors. The results show that our model generated very low multi-step prediction errors, whereas all other methods showed high errors that increased with prediction horizon. 

\label{volunteer_data}

\textcolor{black}{Our final experiment focused on end-to-end volume reconstruction from multiple stack-of-slices with motion parameters estimated by our model and reconstructed with NiftyMIC~\cite{ebner2020automated}. The results of the reconstructions with our estimated motion parameters for 6 subjects in the test set have been shown in Figure S1, and compared favorably with reconstructions using conventional slice-to-volume registration in terms of NRMSE and SSIM. In particular, we achieved average NRMSE of 0.151 with standard deviation of 0.023; and SSIM of 0.912 with standard deviation of 0.031 for our reconstructions. Supplementary Figure S2 shows multi-plane views of a sample case, and Figure S3 shows a case where reconstruction with conventional slice-to-volume registration failed, whereas the reconstruction was improved when we plugged in our estimated motion parameters in the firs iteration of reconstruction.}

\textcolor{black}{The latency for prediction on our hardware (an NVIDIA GeForce 1080 Ti) was $\sim1.42$ \textcolor{black}{ms} per data point where each sequence comprised of 10 slices and outputs were 10 estimations and 10 predictions. Considering the slice acquisition time of $\sim1.5$ seconds for T2-weighted MRI and $\sim80$ \textcolor{black}{ms} for echo-planar imaging, this is real-time.}

\section{Discussion}
\label{sec:discussion}

\textcolor{black}{To the best of our knowledge, this is the first report of the development of a real-time predictive motion tracking technique for fetal MRI. Up until now motion correction in fetal MRI has been done retrospectively through non-causal iterations of slice-to-volume registration and reference volume reconstruction. Slice-to-volume registration is intrinsically an ill-posed problem~\cite{ferrante2017slice}. To overcome this issue, retrospective fetal MRI motion estimation methods that relied on slice-to-volume registration,} evolved from hierarchical~\cite{rousseau2006registration,jiang2007mri} and slice intersection-based~\cite{kim2010intersection} methods to progressive~\cite{gholipour2010robust,kuklisova2012reconstruction,kainz2015fast,ebner2018automated}, patch-based~\cite{alansary2017pvr}, and more recently, dynamic motion estimation techniques~\cite{marami2017temporal}. \textcolor{black}{There have also been nonrigid and deformable extensions of slice-to-volume registration~\cite{uus2020deformable}}.

\textcolor{black}{General-purpose, image-based, MRI motion tracking techniques sought} regularization through modeling motion dynamics~\cite{white2010promo}, or used robust state space models to estimate relative position of sequentially-acquired slices~\cite{marami2016motion,marami2019motion}. While the underlying phenomena are nonlinear, these techniques made simplifying assumptions to linearize the problem and used image registration along with state space estimation by Kalman filtering (or its robust extensions) for motion tracking. Bayesian filtering based Kalman filters fail to model nonlinear relationships as well as non-Gaussian noise, and their extended versions also fail when dynamics are highly nonlinear. These techniques are thus difficult to scale up to real life scenarios\textcolor{black}{, in particular in challenging applications such as fetal MRI}.

More capable Gaussian mixture models~\cite{Ammoun2009RealTT}, process models~\cite{Wiest2012ProbabilisticTP}, or dynamic Bayesian networks (DBN)~\cite{Gindele2015LearningDB} can accommodate complex dynamics but need strong priors by experts which makes them prone to the same practical issues that exist in conventional methods especially when long term prediction is desired. As a result of using image registration, these techniques are computationally intensive and cannot be easily applied in real-time. More importantly, none of the current techniques explicitly uses image information and image recognition to model motion dynamics for 3D pose estimation. Registration-based methods are slow and offer a limited capture range, which makes them prone to failure when motion is continuous and large. In other words, even when integrated with state space estimation methods for dynamic motion tracking, registration-based techniques may not easily recover if they loose subject's position. This is especially problematic in motion estimation in fetal MRI as fetuses in the second and early third trimesters move frequently and rotate in large angles. 
Finally, almost all of the current methods rely on certain initialization assumptions such as the existence of a motion-free reference scan for registration, which is restrictive and unrealistic when considered for use in real-time applications\textcolor{black}{, such as motion tracking for real-time navigation}.

In this paper we showed predictive potential of recurrent neural networks for modeling end-to-end motion in MRI. To this end, we developed a combination of spatial encoders based on convolutional neural networks and temporal encoder and decoder networks based on CNN-LSTM to learn the spatiotemporal features of anatomy and slice sampling from imaging data to predict motion trajectories. Loss functions on multiple regression heads led to a robust model that generalized well beyond the training set to fetuses scanned at different ages and with motion patterns \textcolor{black}{that were recorded from volunteers, which were characteristically different from the synthetic motion patterns that were used in training.} 

\textcolor{black}{To resemble fetal head motion, our volunteer subjects moved their head at different speeds and in different directions to the largest possible extents while we recorded their motion. Comparing the results shown in Figure~\ref{fig:volunter_prediction} (for the fetal test data with recorded motion) with the results in Figure~\ref{fig:pred_compare} (for the fetal test data with synthetic motion) indicates that the average prediction error on recorded data was lower than the average prediction error on synthetic data, despite the fact that the synthetic motion was generated by the same procedure that generated motion patterns in the training data. We attribute this to the fact that the recorded motion had constraints imposed by the mechanical linkage between head and neck that made it easier to predict compared to the synthetic motion.} 

\textcolor{black}{Our approach is a learning-based technique, so its performance depends on what it learns from the training data. Our training data involved large rotations in the -90\degree ~to 90\degree ~range over 15 seconds (10 timesteps). Our training data generation methodology differed from those in earlier 3D pose estimation works, e.g., \cite{salehi2018real2} \cite{hou2017predicting}, which randomly rotated individual slices without taking surrounding slices into account. We generated sequences of interleaved slices covering the 3D anatomy while the anatomy moved on a motion trajectory synthesized by B-Spline curve fitting. This is more realistic than moving slices independently. Yet our model may benefit from training with more realistic simulations of motion, for example using bio-mechanical models of fetal motion~\cite{verbruggen2016modeling} or from ground truth motion recorded from adult volunteers. In this study we used recorded motion only for testing. Dynamic predictive motion tracking, as we proposed here, may also be useful to assess normal versus abnormal patterns of fetal movements~\cite{piontelli2014development} from cinematographic MRI (or 4D ultrasound), which, in-turn, may be used to assess fetal motor behavior~\cite{hayat2018neurodevelopmental,hayat2018neuroimaging}}. 

\textcolor{black}{Obtaining ground truth fetal motion is difficult, especially for large ranges of motion. Motion estimates obtained from successful slice-to-volume reconstructions are typically only available (and reliable) for small ranges of motion. Slice to volume reconstruction techniques rely on 1) redundant slice acquisitions, 2) outlier detection and rejection, and 3) robust reconstructions. Therefore, they effectively filter or remove the effect of mis-registered and motion-corrupted slices~\cite{gholipour2010robust,kuklisova2012reconstruction}. The transformations obtained for the remainder of the slices that are effectively used in reconstruction are typically small; and yet may not be sufficiently reliable to be used as ground truth. Therefore, to test our approach on original fetal MRI scans with motion, we used our estimated motion parameters along with a powerful slice-to-volume reconstruction method~\cite{ebner2020automated} to reconstruct volumes from multiple stack-of-slices. Reconstruction with our estimated motion parameters compared favorably against reconstruction with retrospective slice-to-volume registration (Supplementary Figures S1-S3).}

\textcolor{black}{Our model generalized well to data from subjects at ages outside of the age range of the training data and with realistic motion patterns that were never used in training. Initially we found that the model had difficulty estimating motion for large and fast movements. To resolve this we used curriculum training which trained the network on difficult samples more often than easier ones that alleviated the issue. Our initially trained models also had difficulty generalizing to unseen gestational ages with large and fast movement in the validation set. To resolve that, we added batch normalization and regularized by reducing the number of parameters in the model which resulted in better performance. Yet, since our method is a learning-based approach, its performance is expected to degrade if there is significant domain shift between the training and test data. For example, the performance of our model may significantly drop if a different modality or sequence is used as test, or if a significantly different set of parameters are used in fetal MRI scans. To adapt the model to new domains, domain adaptation techniques or pre-processing may be used, e.g.~\cite{salehi2018real2}. Also, our trained model may not generalize well for tracking motion of severely abnormal anatomies. Possible remedies for this problem are to include abnormal cases in training and to use curriculum learning with appropriate data augmentation. These are excellent directions for future work.}

\textcolor{black}{Our model architecture is small compared to most state-of-the-art RNNs. This helped us achieve real-time performance. Curriculum training helped the network focus on more difficult samples, i.e. sequences with large and fast motion. We kept our model a causal predictive model for its intended application which is real-time motion tracking and navigation. For other applications, such as retrospective processing of image time series, using signal from the future, e.g. by bidirectional LSTM, is expected to increase performance but would break the causal nature of the model. To train our model we used the MSE loss due to its well-posed convex nature for optimization in our high-dimensional search space. For static pose estimation~\cite{salehi2018real2}, a second stage optimization (refinement) with the geodesic loss, which is a natural Riemannian metric on the compact Lie group $SO(3)$ of orientations, improved the results.} 
\textcolor{black}{We observed a similar trend here but at a relatively lower degree. By fine tuning our model (trained with MSE) for 10 additional epochs at a learning rate of 0.0001 with geodesic loss we observed average error reduction of 0.4\degree ~in estimation, which was statistically significant; but did not see a statistically significant reduction in prediction error.} 




\textcolor{black}{By observing a sequence of slices, our trained model predicts the relative 3D pose (motion) of the anatomy with respect to an initial pose. For real-time slice navigation, therefore, we require an estimate of the initial pose; which can be obtained by the pose estimation techniques proposed in~\cite{hou20183d} and \cite{salehi2018real2}. Although those techniques can accurately estimate the 3D pose of the fetal brain in a canonical (atlas) space based on a volume or a slice (or stack of slices) close to the center of the anatomy, their estimation error is relatively high in the border slices where image features are sparse, and their predictive performance is limited for fast and large motion. Experimental results in motion tracking showed that our technique outperformed time series prediction models built upon those static pose estimation methods. Therefore, to build an effective and efficient real-time fetal MRI navigation system, a combination of initial pose estimation by techniques such as those proposed in~\cite{hou20183d} and \cite{salehi2018real2} and our predictive motion tracking technique is needed. Echo-planar imaging~\cite{afacan2019fetal} may be an appropriate choice to acquire fast volumes (as 3D localizer or navigator) to estimate initial pose at the beginning or in intervals between sequences.} 

\section{Conclusion}
\label{sec:conclusion}

\noindent \textcolor{black}{
We developed and presented a technique that is capable of estimating and predicting the 3D pose trajectories of the fetal brain in real-time despite large fetal movements. This technique, when augmented with other real-time components and implemented on MRI scanner platforms, may be used to track fetal head motion as slices are acquired, make recommendations for scan orientations as a decision support system or a human-in-the-loop navigation system, or to build real-time automatic fetal MRI systems, which, in-turn, can lead to much more efficient, effective, and tolerable fetal MRI scan sessions. Real-time predictive motion tracking can also play a critical role in real-time assessment of the quality of highly motion-sensitive scans such as fetal functional MRI that are very difficult to perform, and to automatically adapt the duration of such scans to ensure data of sufficient quality is acquired for post-acquisition processing. Finally, image-based dynamic motion tracking can also be used to assess fetal movements and motor behavior \textit{in-utero} from cine MRI and 4D ultrasound.}




\ifCLASSOPTIONcaptionsoff
  \newpage
\fi

\bibliographystyle{IEEEtran}
\bibliography{main}

\end{document}


\maketitle

\textbf{Figure S1}: Volume reconstruction from multiplanar fetal MRI scans with motion parameters estimated with our method (Our) compared to reconstruction with slice-to-volume registration (SVR) for 6 fetuses in the test set. The normalized root mean square error (NRMSE) (lower is better) and structural similarity image metric (SSIM) (higher is better) have been reported between the two reconstructions for each subject. These results show that retrospective volume reconstructions with the motion parameters estimated using our method compared favorably against retrospective SVR.\\

\textbf{Figure S2}: Three plane views (coronal, sagittal, and axial) of volumes reconstructed by NiftyMIC using motion parameters estimated with our method (Our, top) and with slice-to-volume registration (SVR, middle), and the error between the two reconstructions (bottom).\\

\textbf{Figure S3}: An example of a failed volume reconstruction using slice-to-volume registration (SVR, bottom), where our method improved the reconstruction (Our, top).\\

\begin{figure*}[h!]
    \centering
    \includegraphics[keepaspectratio, width=0.85\textwidth]{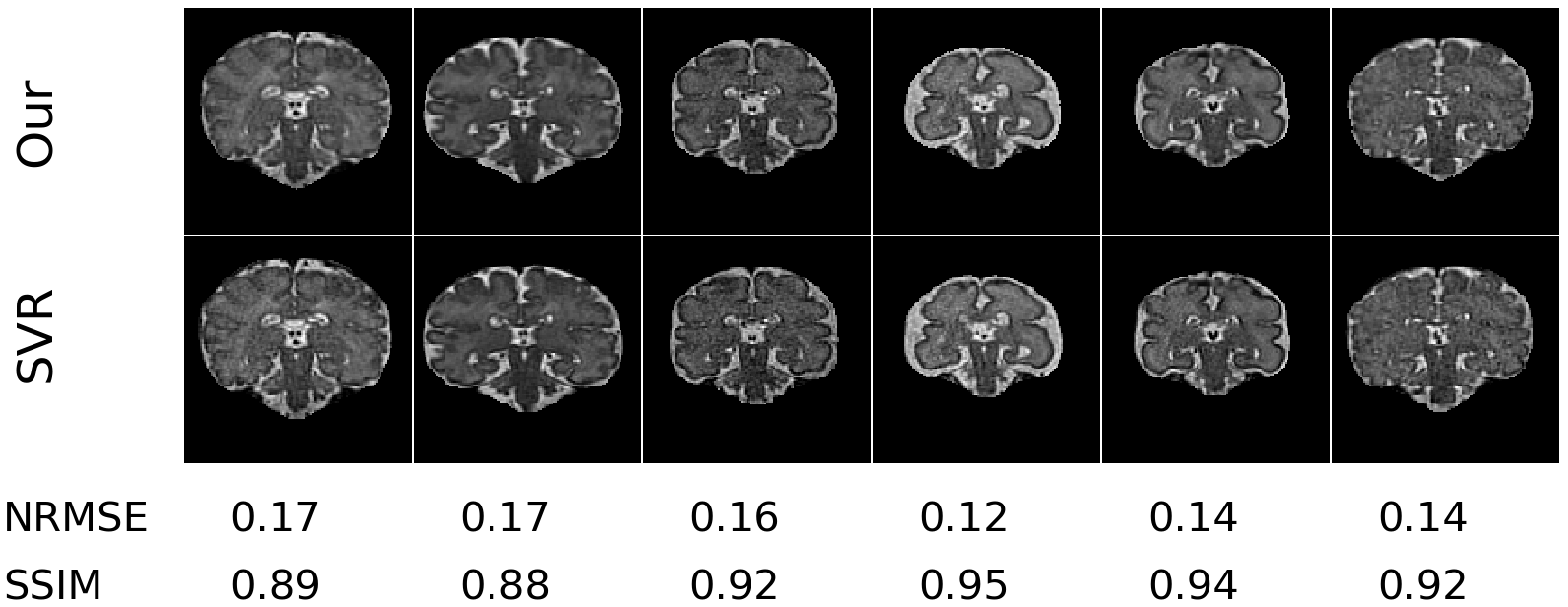}
    \caption{}
    \label{fig:boxplot_nrmse_ssim}
\end{figure*}



\begin{figure}[]
\begin{minipage}{.5\textwidth}
  \centering
  \includegraphics[width=0.78\linewidth]{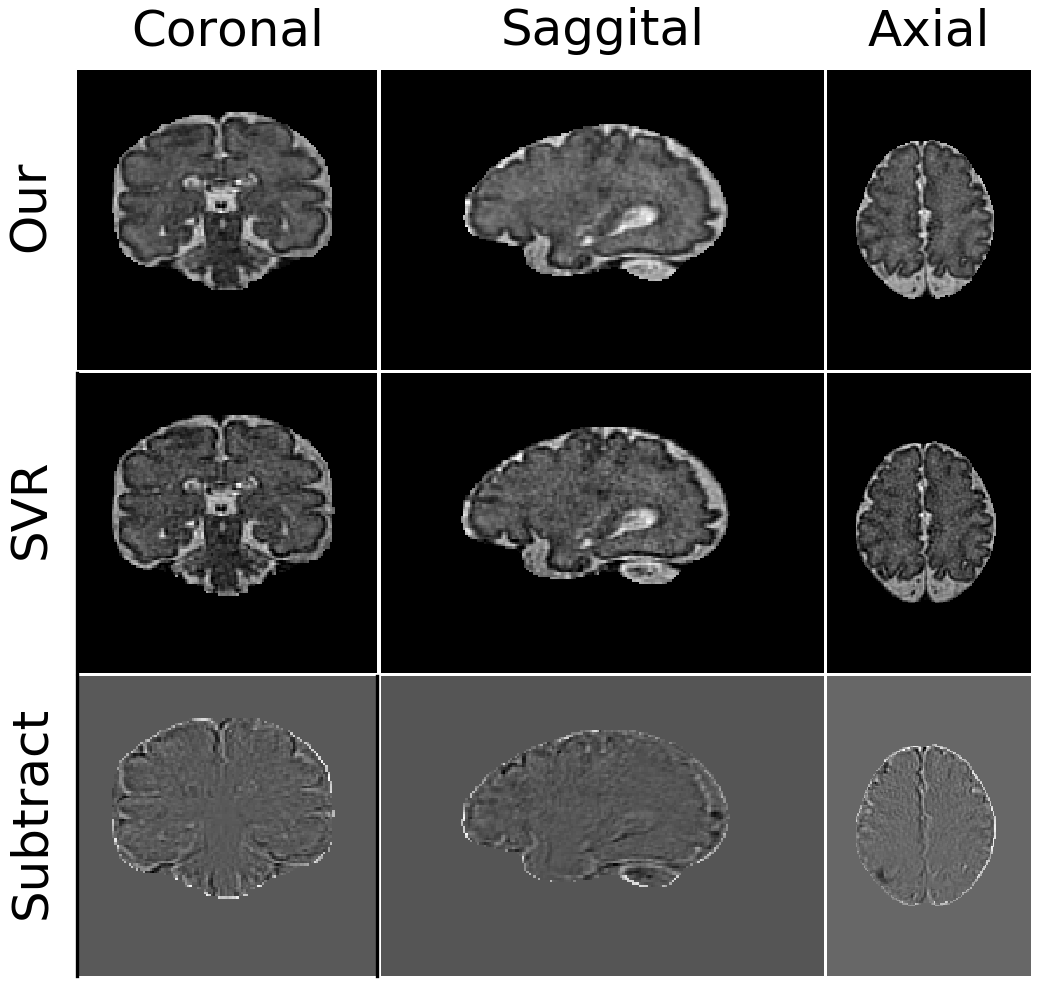}
  \caption{}
  \label{fig:sub-second}
\end{minipage}
\begin{minipage}{.5\textwidth}
  \centering
  \includegraphics[width=1.0\linewidth]{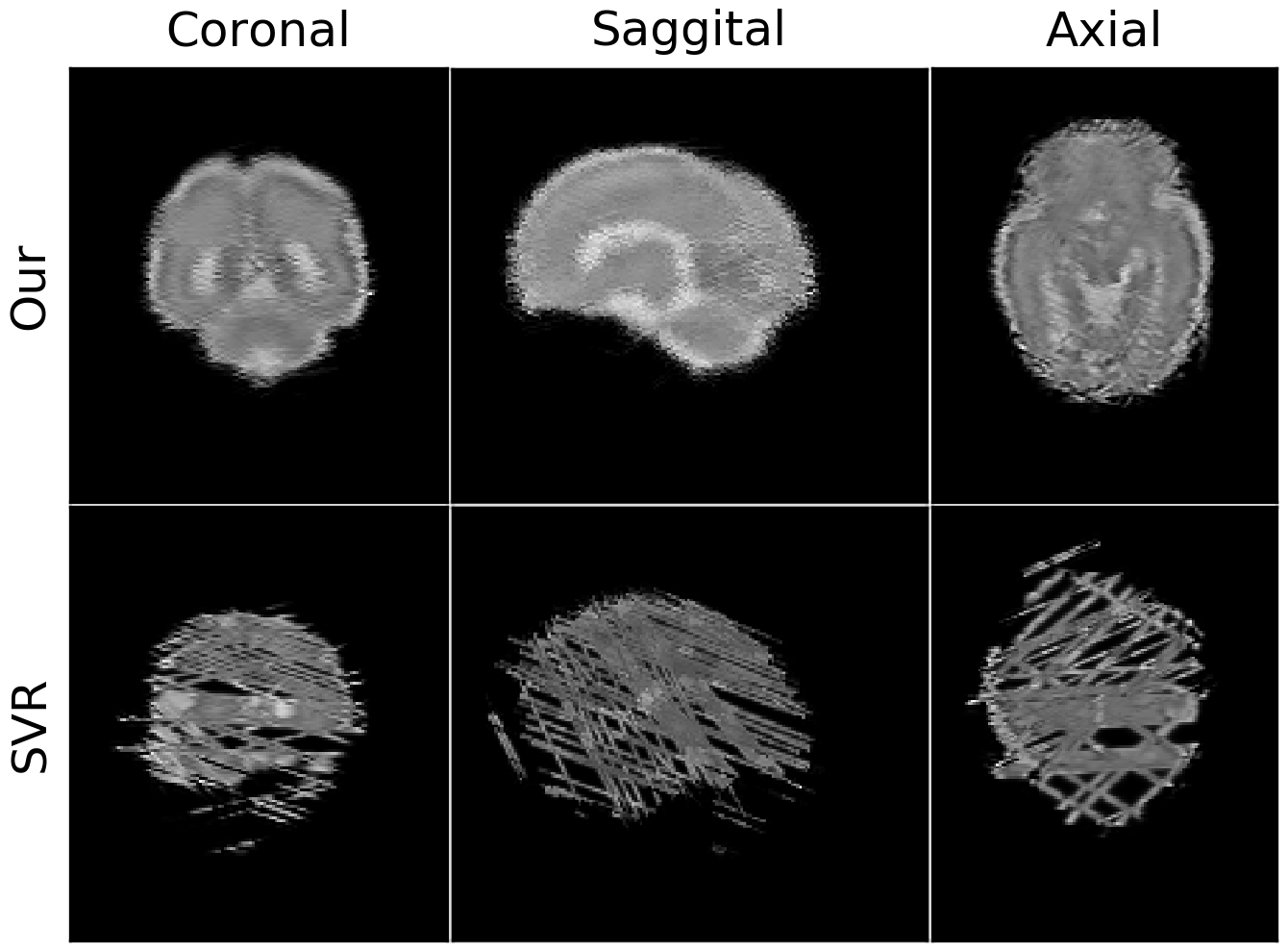}  
  \caption{}
  \label{fig:sub-first}
\end{minipage}
\label{fig:fig}
\end{figure}



\ifCLASSOPTIONcaptionsoff
  \newpage
\fi